\def\r{\rho}
\def\o{\omega}
\def\beg{\begin{equation}}
\def\eeq{\end{equation}}
\documentstyle[12pt]{article}
\begin{document}
\begin{center}
{\Large{\bf Illumination of Quantum Hall States}}
\vskip0.35cm
{\bf Keshav N. Shrivastava}
\vskip0.25cm
{\it School of Physics, University of Hyderabad,\\
Hyderabad  500046, India}
\end{center}

We show that illumination by continuous white radiation saturates
the population of levels so that some of the transitions are not
observed. This reduces the number of observable transitions so
that the resolution of neighboring transitions is improved.
In the case of $\nu=11/2$ the improved resolution leads to clear observation of zero in the $\r_{xx}$ which appears as a finite resistivity minimum in the absence of saturating radiation.
\vfill
Corresponding author: keshav@mailaps.org\\
Fax: +91-40-3010145.Phone: 3010811.
\newpage
\baselineskip22pt
\noindent {\bf 1.~ Introduction}

In ordinary spectroscopy, in some cases, the absorption lines
are so wide that lines overlap resulting into a poorly resolved
spectrum. If the system is illuminated by a pulse of radiation
of a resonant frequency, the population of the upper level can
be saturated so that the line corresponding to this frequency
does not absorb and in fact an emission occurs from the
saturated level so that there appears a hole in the absorption
spectrum. This type of hole burning in the absorption spectrum
is well known and a systematic study of this effect was
performed by Khutsishvili et al [1-5]. If we illuminate the
system with a white light, absorption can occur at many levels
and the population can be disturbed. The line shape clearly
shows the frequencies at which saturation has occured. At this
time, the detailed theories of ``fractional charge" are  believed to be ``not applicable" to the actual experimental data.

In this letter, we propose an interpretation of quantum Hall
resistance in terms of energy levels. It then follows that a
plateau should occur at $\nu H$ with $\nu=11/2$. The magnetic
field $H$ fixed by the value $\nu H$ also describes a zero
value in the $\r_{xx}$. This means that the resistivity is
highly anisotropic in the $xy$ plane. By disturbing the
population of neighboring levels, we can unmask the zero value
of $\r_{xx}$ at $\nu=11/2$, i.e., the zero value of $\r_{xx}$ is
not found at $\nu=11/2$ when the sample is dark but only a
finite value occurs. When illumination by a red light is turned
on, there is absorption at the neighboring levels and the zero
value at $\nu=11/2$ becomes visible.
\vskip0.25cm
\noindent{\bf 2.~~Theory}

In our theory [6] the effective charge arises from the effective
value of the Bohr magneton, $\mu_B=e\hbar/2mc$. The values of
the orbital angular momenta $l$ and the spin $s$ are combined in
such a way that an effective charge can be defined,
\beg
e_{eff} = {1\over2} ge = \nu_\pm e
\eeq
where
\beg
\nu_+ = {l+1\over2l+1}
\eeq
and
\beg
\nu_- = {l\over2l+1}\,\,\,.
\eeq
For $l=0$, we obtain, $\nu_+=1$ and $\nu_-=0$ and for $l=1$, we
obtain $\nu_+=2/3$ and $\nu_-=1/3$ which describe the effective
charge. The values of the effective charges which we tabulate
are the same as those experimentally found. The subscript $-$ in
$\nu_-$ indicates that spin is $-{1\over2}$ and that in $\nu_+$
shows that spin is $+{1\over2}$. Usually $\hbar\omega=g\mu_BH$
gives the resonance. However, in the present case
$\hbar\omega=\nu_{\pm}\mu_BH$. We consider that this frequency
is a cyclotron frequency and transition with $n\hbar\omega_c$
are observable. Therefore, we can multiply the values of
$\nu_\pm$ by an integer. Of course, this number, $n$, can be
identified as the Landau level quantum number. Therefore,
transitions at $2\o_c$, $3\o_c$, etc. occur. With $l=\infty$,
both of the above series give $\nu_+(l\to\infty)={1\over2}$ and
$\nu_-(l\to\infty)={1\over2}$. Since we can multiply these
numbers by an integer, we predict the observable frequencies as
$n/2$. For $n=11$, we obtain (11/2). As the value of $n$
increases, the intensity of the line reduces and hence for
$n=11$ the transition becomes considerably weak. Now, we have to
convert the absorption response to the resistivity. This
conversion is straightforward [7]. So far we have obtained the
correct effective charges, particle-hole symmetry [8] and doubly
degenerate state at $n/2$. What is found as energy levels in the
absorption becomes plateaus in the transverse resistivity,
$\r_{xy}$ as a function of magnetic field. At a field slightly
higher than that at 11/2=5.5, occurs 16$\times$(1/3)=16/3=5.33.
If 16/3 is saturated by illumination of light, the transition at
16/3 disappears and the resolution near 11/2 improves. The
improvement can be so good that while the zero of 11/2 is not
clearly visible it becomes visible when the sample is
illuminated by a continuous wave light.
\vskip0.25cm
\noindent{\bf 3.~~Experimental data}

Cooper et al [9] have performed the measurements of longitudinal
resistance, $\r_{xx}$, along [1,$\bar{1}$,0] and along [110]
directions, as a function of magnetic field. At low fields, such
as $2T$ very weak structure in the resistivity has been detected
and the minima in $\r_{xx}$ corresponding to plateaus in
$\r_{xy}$ have been detected at $\nu=9/2$ and 11/2. The minimum
at $\nu=11/2$ is quite clearly seen. The sample is then
illuminated by a red light emitting diode. The level saturation
effects are observed, the same way as predicted by us using the
energy level representation for the quantum Hall effect. Due to
the saturation at 16/3, the zero value at 11/2 becomes quite
clear. Therefore, the experimental data is in agreement with
what is theoretically predicted on the basis of energy levels.
\vskip0.25cm
\noindent{\bf 4.~~ Interactions}

We see that 1/3 comes from $s=-1/2$ whereas 2/3 comes from
$+{1\over2}$. Therefore, there is a need of an interaction which
can flip the spin when magnetic field is varied. The interaction
can be of the form ${\bf l.s}=l_xs_x+l_ys_y+l_zs_z$ so that both $l$
as well as $s$ have to flip. However, for going from 1/3 to 2/3,
$l$ need not change. Therefore, the interaction is of the form
$l_zs_x$ which can happen in the case of a very anisotropic
interaction. Accordingly, we can write the interaction as
\beg
{\cal H}^\prime = \sum_i \lambda_i<l^i_z>(s^i_++s^i_-)
\eeq
where the summation is over all of the electrons. Next we
consider as to how we can go from one value of $l$ to another
value of $l$. This can be done by two site exchange
interaction $l_+^il_-^j$ with pairwise summation,
\beg
{\cal H}^\prime = \sum_{i>j}^\prime j_{ij}(l^i_+l_-^j + l_+^j l_-^i)\,\,\,.
\eeq
It seems that the above interaction, (4) is quite
sufficient to produce experimentally observed phenomenon and
we need an $L_\pm$ operator which should come from some where but not
necessarily from (5) above, which causes an exchange interaction. The values given by ref. [6] are the same as those found by St\"ormer [10]
and hence if $\bf J$ is treated properly with both signs in ${\bf J}$ = ${\bf L\pm S}$ no interactions are needed to locate the plateaus except the shift operator. However the width of the plateaus obviously requires the many-body electron-phonon type interactions.

\noindent{\bf 5.~~Flux Tubes and Composite Fermions}

The composite fermion theory requires that even number of fluxes
are attached to the electron. The data of Cooper et al does not
show any evidence of flux quantization with even number of
fluxes. Therefore, it is clear that composite fermion theory is
not in conformity with the experimental data. In 1989, Jain has
suggested [11] that even number of fluxes are attached to the
electron so that the magnetic field becomes, $B^*=B-2\rho\phi_o$
where $\rho$ is the density of electrons. The factor of 2 is
used so that only even number of flux quanta, $2\phi_o$ in this
case, are used. The experimental data does not agree with this
``even number'' quantization of magnetic field shift. There is
no evidence of even number multiplied to any quantity in the
experimental data. We have found [12] that the Jain's theory of
composite fermions is internally inconsistent. Therefore, it is
concluded that flux tubes are not attached to the electrons.
\vskip0.25cm
\noindent{\bf 6.~~Lande's $g$-value formula}

The Lande's formula for the $g$-value of the electron in atomic
physics gives only one value given in terms of $L$ and $S$ of
the atom as,
$$
g = 1 + {J(J+1)-L(L+1)+S(S+1)\over2J(J+1)}\,\,\,.
$$
For ${\bf J=L+S}$, the values of $g/2$ for various values of $L$ and
$S={1\over2}$, are given by
\vskip0.15cm
\begin{center}
\begin{tabular}{cccccc}
$L$     & 0 & 1   & 2   & 3 \\
$g_-/2$ & 1 & 2/3 & 3/5 & 4/7& etc.
\end{tabular}
\end{center}
\vskip0.15cm
For ${\bf J=L-S}$, the above values become,
\vskip0.15cm
\begin{center}
\begin{tabular}{cccccc}
$L$     & 0 & 1   & 2   & 3 \\
$g_+/2$ & 0 & 1/3 & 2/5 & 3/7& etc.
\end{tabular}
\end{center}
\vskip0.15cm
Since the Bohr magneton has the charge of the electron the above
$g_\pm$ give the effective charge. In the case of $L=0$,
$g_{(-)}/2$ has a zero charge solution which gives the soft mode
or symmetry breaking mode. The values tabulated above agree with
those found in the experimental data [10]. Usually, there is
only one value of $g$ for a given value of $L$ which is measured
in the electron spin resonance. In the present case of quantum
Hall effect, $L$ is not a constant and changes as the magnetic
field is varied. The plateaus in the quantum Hall effect therefore
occur at the energy levels determined from the single electron expression of $g$ values and the relaxation times which become widths of the plateaus are determined from the electron-phonon interaction
which is a many-body interaction. 

Our theory also agrees with data at large values of $L$ [13]. We
have compared a lot of experimental data with our theory and
found good agreement in all cases [14,15]. We find that the
magnetic moment of the electron is slightly modified at large
magnetic fields [16]. The polarization of the half-filled level
is also predicted correctly by using Knight shifts [17]. At the
half-filled Landau level, there is a symmetry breaking resulting
into the appearance of a Goldstone boson in bilayers of
semiconductors which are predicted correctly [18].

Modern and elegant theories in which fractional charge can be
understood on the basis of appropriate generalizations of the
Laughlin-type treatment are obviously irrelevant to quantum Hall
effect experiments. The area of the flux quantization as well as
the spin has not been treated correctly by Laughlin and in view
of proper calculations of angular momenta, the need for the
fractionally charged wave function completely disappears.
Schoutens [19-22] has made an effort to consider the spin but
the physics of the problem in his theory is  not
relevant to the experimental work on quantum  Hall effect.
\vskip0.25cm
\noindent{\bf7.~~Conclusions.}

We conclude that the plateau at 11/2 arises from $L\to\infty$
limit in (2) and (3) multiplied by $n=11$ because of
$n\hbar\omega$ transitions. When $n$ becomes large, the
intensity reduces. This predicted reduction in intensity for
large values of $n$ is in agreement with the experimental data
of Cooper et al [9].
\newpage
\noindent{\bf8.~~References}
\begin{enumerate}
\item P.I. Bekauri, B.G. Berulava, T.I. Sanadze, O.G.
	Khakhanashvili, G.R. Khutsishvili, Sov. Phys. JETP
	{\bf32}, 200 (1971) [Zh. Eks. Teor. Fiz. {\bf59}, 368
	(1970)]. 
\item T.I. Sanadze and G.R. Khutsishvili, Sov. Phys. JETP
	{\bf32}, 412 (1971) [Zh. Eks. Teor. Fiz. {\bf59}, 753
	(1970)]. 
\item N.S. Bendiashvili, L.L. Buishvili and G.R. Khutsishvili,
	Sov. Phys. JETP {\bf30}, 671 (1970). [Zh. Eks. Teor. Fiz.
	{\bf57}, 1231 (1969)].
\item T.I. Sanadze and G.R. Khutsishvili, Sov. Phys. JETP
	{\bf29}, 248 (1969), [Zh. Eks. Teor. Fiz. {\bf56}, 454
	(1969)]. 
\item L.L. Buishvili, M.D. Zviadadze and G.R. Khutsishvili, Sov.
	Phys. JETP {\bf29}, 159 (1969), [Zh. Eks. Teor. Fiz.
	{\bf56}, 290 (1969)].
\item K.N. Shrivastava, Phys. Lett. A{\bf113}, 435 (1986);
	{\bf115}, 459(E)(1986).
\item K.N. Shrivastava, arXiv: cond-mat/0104577 (2001).
\item K.N. Shrivastava, Mod. Phys. Lett. B{\bf13}, 1087 (1999).
\item K.B. Cooper, M.P. Lilly, J.P. Eisenstein, T. Jungwirth,
	L.N. Pfeifler and K.W. West, Solid State Commun.
	{\bf119}, 89 (2001). 
\item H.L. St\"ormer, Rev. Mod. Phys. {\bf71}, 875 (1999).
\item J.K. Jain, Phys. Rev. Lett. {\bf63}, 199 (1989); K. Park
	and J.K. Jain, Phys. Rev. Lett. {\bf81}, 4200 (1998).
\item K.N. Shrivastava, arXiv: cond-mat/0105559 (2001).
\item K.N. Shrivastava, Mod. Phys. Lett. B{\bf14}, 1009 (2000);
	arXiv: cond-mat/0103604 (2001).
\item K.N. Shrivastava, CERN SCAN-0103007 (2001).
\item K.N. Shrivastava, in Frontiers of Fundamental Physics 4,
	edited by B.G. Sidharth and Altaisky, Kluwer Academic/Plenum
	Press, N.Y. 2001.
\item K.N. Shrivastava, arXiv: cond-mat/0104004 (2001).
\item K.N. Shrivastava, arXiv: cond-mat/0106160 (2001).
\item K.N. Shrivastava, arXiv: cond-mat/0106445 (2001).
\item E. Ardonne and K. Schoutens, Phys. Rev. Lett. {\bf82},
	5096 (1999).
\item E. Ardonne, P. Bouwknegt, S. Guruswamy and K. Schoutens,
	Phys. Rev. B{\bf61}, 10298 (2000).
\item K. Schoutens, Phys. Rev. Lett. {\bf81}, 1929 (1998).
\item P. Bouwknegt and K. Schoutens, Phys. Rev. Lett. {\bf82},
	2757 (1999).
\end{enumerate}
\end{document}